\documentclass[10pt,a4paper,twocolumn,english,prl,aps,showpacs,floatfix,groupedaddress,superscriptaddress]{revtex4}
\usepackage[T1]{fontenc}
\usepackage[latin1]{inputenc}
\usepackage{amsmath}
\usepackage{graphicx}
\usepackage{amssymb}

\makeatletter
\newcommand{\tr}{\mathrm{tr}}

\def\1{1\negthickspace{\rm I}}

\usepackage{babel}
\makeatother
\begin{document}

\title{Long-distance entanglement in spin systems }

\author{L. Campos Venuti}

\affiliation{Dipartimento di Fisica, V.le C. Berti-Pichat 6/2, I-40127 Bologna,
Italy}

\affiliation{INFN Sezione di Bologna, V.le C. Berti-Pichat 6/2, I-40127 Bologna,
Italy}

\author{C. Degli Esposti Boschi}

\affiliation{Dipartimento di Fisica, V.le C. Berti-Pichat 6/2, I-40127 Bologna,
Italy}

\affiliation{CNR-INFM, Unità di Bologna, V.le C. Berti-Pichat 6/2, I-40127 Bologna,
Italy}

\author{M. Roncaglia}

\affiliation{Dipartimento di Fisica, V.le C. Berti-Pichat 6/2, I-40127 Bologna,
Italy}

\affiliation{INFN Sezione di Bologna, V.le C. Berti-Pichat 6/2, I-40127 Bologna,
Italy}

\affiliation{CNR-INFM, Unità di Bologna, V.le C. Berti-Pichat 6/2, I-40127 Bologna,
Italy}

\date{\today}

\begin{abstract}
Most quantum system with short-ranged interactions show a fast decay
of entanglement with the distance. In this Letter, we focus on the
peculiarity of some systems to distribute entanglement between distant
parties. Even in realistic models, like the spin-1 Heisenberg chain,
sizable entanglement is present between arbitrarily distant particles.
We show that long-distance entanglement appears for values of the
microscopic parameters which do not coincide with known quantum critical
points, hence signaling a transition detected only by genuine quantum
correlations. 
\end{abstract}

\pacs{03.67.Mn, 75.10.Pq, 05.70.Jk}

\maketitle
Entanglement generation and distribution is a problem of central importance
in performing quantum information (QI) tasks, like teleportation \cite{boschi98}
and quantum cryptography \cite{gisin02}. Typically, the entanglement
between parties is created by means of a direct interaction. Since
entanglement needs the presence of strong correlations, low dimensional
systems, as for example antiferromagnetic spin chains, offer a natural
source of entanglement. In most systems with short-range interactions,
the entanglement between a pair of particles decays rapidly with the
distance (generally even more rapidly than standard correlations).
For example, in the Ising model with transverse field \cite{osborne02,osterloh02}
the concurrence vanishes for distances larger than two sites, while
in the Heisenberg model \cite{takahashi77,jin03} it is restricted
only to nearest neighbors. 

From the QI perspective, it would be attracting to create sizable
entanglement between particles that are located at a distance larger
than a few sites. Along this direction, the localizable entanglement
(LE) was conceived with the idea of exploiting spin chains as quantum
channels \cite{verstraete04a,campos05}. The LE measures the average
entanglement localized between a couple of distant points, after performing
optimal local measurements onto the rest. 

In the present Letter, we show that already the ground state (GS)
of various models widely used in condensed matter physics offer the
possibility to entangle parties that are arbitrarily far apart. This
fact, naturally leads to the concept of long distance entanglement
(LDE) as a sort of quantum order parameter. As discussed in the following,
the onset of LDE does not coincide with known quantum phase transitions
(QPT's) of the systems we have examined. 

Let us consider two sites A and B that interact with a many-body system
C. The distance $d$ between A and B is set by the individual short-ranged
interactions in the subsystem C (see figure below). 

\begin{center}\includegraphics[%
  scale=0.3]{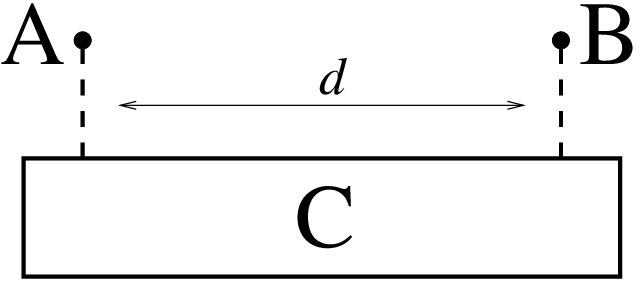} \end{center}

According to our definition, given a bipartite measure of entanglement
$E(\rho)$, we have LDE if 

\[
E_{d}\left(\rho_{\textrm{AB}}\right)\stackrel{d\to\infty}{\longrightarrow}E_{\infty}\neq0\]
where $\rho_{\textrm{AB}}={\rm Tr}_{\textrm{C}}\left|\Psi\right\rangle \left\langle \Psi\right|$
is the reduced density matrix of the subsystem A and B, and $\left|\Psi\right\rangle $
is the total wavefunction. The introduction of two special points,
or probes, is essential here since the property of monogamy \cite{coffman00}
limits to two the number of particles maximally entangled. The basic
idea comes from the observation that if we wish to locate a great
amount of entanglement between two selected qubits, we are forced
to exclude entanglement with the rest. Specifically, we have considered
cases where A and B represent end spins in an open chain or additional
spins (probes) that interact with selected sites in the chain. In
condensed matter systems, these might be impurities, defects or even
scattering particles \cite{dechiara05}. 

As a first criterion, we expect to have a nonvanishing LDE between
A and B when their interactions with C are \emph{small} compared to
the typical interactions contained in C. Otherwise, A or B would develop
too strong correlations with the closest degrees of freedom in C,
excluding the possibility to form LDE. On the other side, strong correlations
among the particles in C will tend to avoid entanglement between C
and the probes. In this sense, strongly correlated quantum systems,
like antiferromagnetic spin systems, are good candidates to do the
job. In particular, spin-1/2 antiferromagnetic systems admit a simple
picture based on resonating valence bonds (RVB) \cite{Anderson87}.
If a state is a total spin singlet, then it may be approximated by
all the possible RVB configurations, each one with a given weight.
Resonances between various configurations destroy entanglement. The
variational idea for favoring a singlet between two selected sites
(A and B) is to induce a large weight for all the RVB configurations
that link pair of particles inside C by increasing here the interactions. 

In the following, we will present some mechanisms able to produce
LDE in spin-1/2 and spin-1 chains.

\paragraph{The dimerized-frustrated model. }

In the $S=1/2$ antiferromagnetic isotropic Heisenberg chain, each
spin is highly entangled with its nearest-neighbors \cite{wootters01}.
Instead we consider here the dimerized chain with frustration, well-known
for its connections with spin-Peierls \cite{augier97} and ladder
compounds, whose Hamiltonian is \begin{equation}
\mathcal{H}=\sum_{j=1}^{L-1}\left[1+\delta(-1)^{j}\right]\vec{\sigma}_{j}\cdot\vec{\sigma}_{j+1}+\alpha\sum_{j=1}^{L-2}\vec{\sigma}_{j}\cdot\vec{\sigma}_{j+2}\label{eq:alpha_delta_model}\end{equation}
where $\sigma^{\nu}$, $\nu=x,y,z$ are Pauli matrices. For $\delta=0$
the system is gapless up to $\alpha_{c}\approx0.241$, where the GS
spontaneously dimerizes and becomes doubly degenerate. In the Majumdar-Ghosh
line $\delta+2\alpha=1$ the system is made only by short-ranged singlets. 

We choose $L$ even and open boundary conditions (OBC), with the aim
to study the entanglement between the two spin-$1/2$ at the end points.
First, let us look at two limit cases. For $\delta=-1$ and $\alpha=0$
the GS is dimerized onto the ``odd'' bonds,

\begin{center}\includegraphics[%
  scale=0.3]{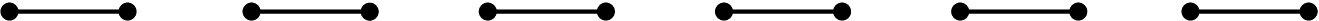} \end{center}

\noindent where the entanglement is localized in pairs of nearest-neighbors.
More interesting for us is the case $\delta=1$ and $\alpha=0$, where
two spins are ``left alone'' as in the following figure

\begin{center}\includegraphics[%
  scale=0.3]{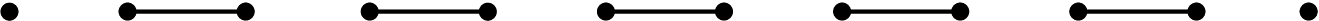}\end{center}

\noindent and the GS is fourfold degenerate. The basic idea for concentrating
a large amount of entanglement between end spins (A and B) is related
to their tendency to form a global singlet in the GS for any $\delta\not=1$.
Hence the two end states are forced to develop strong correlations
towards the formation of a long-distance singlet state $\left|\Psi^{-}\right\rangle \equiv\left(\left|\uparrow\downarrow\right\rangle -\left|\downarrow\uparrow\right\rangle \right)/\sqrt{2}$.
This phenomenon can be thought of as a long range antiferromagnetic
interaction mediated by the other spins in the chain. Accordingly,
the states in $S_{tot}=1$ form a triplet of excitations. 

As a measure of entanglement, we adopt the concurrence \cite{wootters98}.
Given the SU(2) invariance of the GS, the spin-spin correlations $\gamma_{ij}^{\nu\nu}\equiv\left\langle \sigma_{i}^{\nu}\sigma_{j}^{\nu}\right\rangle /4$
are the same for every $\nu$. In addition, the magnetization is zero,
so that the concurrence between A and B reduces simply to $\mathcal{C}_{\textrm{AB}}=2\max\left\{ 0,2\left|\gamma_{\textrm{AB}}^{zz}\right|-\gamma_{\textrm{AB}}^{zz}-1/4\right\} $.
The concurrence is nonzero if the antiferromagnetic correlations between
A and B are sufficiently strong: $\gamma_{\textrm{AB}}^{zz}<-1/12$. 

\begin{figure}
\begin{center}\includegraphics[%
  scale=0.45]{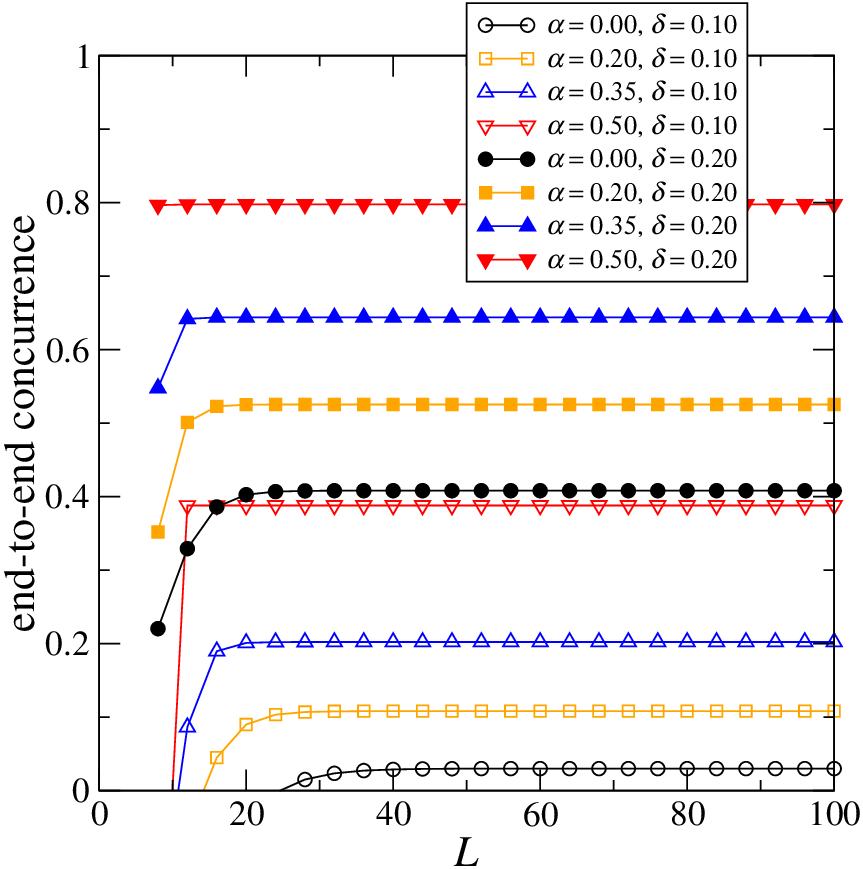}\end{center}

\caption{The finite size study on the concurrence shows the presence of LDE
in model (\ref{eq:alpha_delta_model}) for $\delta>\delta_{T}(\alpha)$.
Data were obtained keeping 256 DMRG states, with a truncation error
smaller than $10^{-10}$. \label{cap:finite-size}}
\end{figure}

First, we have performed some numerical evaluations on the GS using
the density matrix renormalization group (DMRG) method \cite{schollwock05}.
The end-to-end concurrence $\mathcal{C}_{\textrm{AB}}$ is plotted
in Fig. \ref{cap:finite-size} as a function of the system size $L$
for several values of $\delta$ and $\alpha$. The numerical data
put in evidence the presence of LDE as well as the rapid achievement
of the asymptotic value. This latter feature is consistent with the
small correlation length in the regime $\delta\gtrsim0.10$ (see e.g.~\cite{augier97})
and allows us to study shorter chains by means of an exact diagonalization
program based on the Lanczos method. 

\begin{figure}
\begin{center}\includegraphics[%
  scale=0.45]{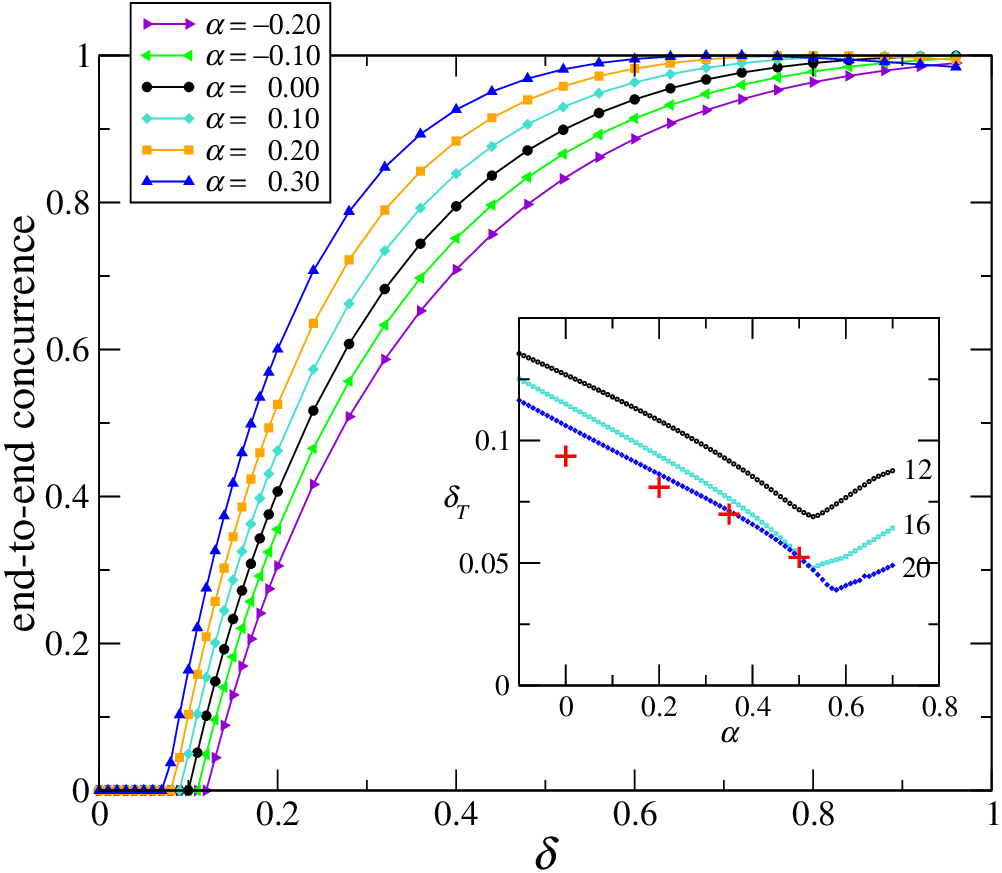}\end{center}

\caption{Concurrence for the two end-qubit state as a function of dimerization
$\delta$ for some values of $\alpha$. Exact calculation on a chain
of length $L=24$. The LDE increases steeply above a threshold and
is enhanced by frustration. Inset: threshold value of dimerization
$\delta_{T}(\alpha)$, above which end-to-end concurrence starts to
be nonzero for lengths $L=12,16,20$. The crosses are the infinite
size extrapolations of DMRG data with $L$ up to 100. \label{cap:Concurrence vs delta}}
\end{figure}

The numerical results summarized in Fig.~\ref{cap:Concurrence vs delta}
shows that the end-to-end concurrence grows rapidly with $\delta$
starting from a threshold value $\delta_{T}(\alpha)$. For $\delta=0$
no LDE is generated, and this is related to the tendency of the first
spin to entangle with the second, as found in Ref. \cite{Affleck06}.
This is also consistent with the absence of surface order in an open
$S=1/2$ Heisenberg chain \cite{alcaraz88}.

\noindent %

The inclusion of $\alpha<0$ tends to favor a classical N\'eel state,
so it is expected to destroy entanglement. On the contrary, $\alpha>0$
is seen to enhance the end-qubit concurrence, as frustration favors
quantum fluctuations. This is shown in the inset of Fig.~\ref{cap:Concurrence vs delta}
where $\delta_{T}(\alpha)$ decreases with $\alpha$, reaching a minimum
for $\alpha\approx0.5$. In the limit $\left|\alpha\right|\gg1$ the
entanglement gets suppressed as the probes belong to two separated
chains. Remarkably, from Figs.~\ref{cap:finite-size} and \ref{cap:Concurrence vs delta}
it emerges that the entanglement grows with the system size $L$.

\paragraph{Spin-1 Chain. }

An important class of spin-1 models is given by the Heisenberg chain
with biquadratic interactions, 

\[
\mathcal{H}=\sum_{i=1}^{L-1}\left[\vec{S}_{i}\cdot\vec{S}_{i+1}+\beta\left(\vec{S}_{i}\cdot\vec{S}_{i+1}\right)^{2}\right]\]
that has attracted much interest both for the study of hidden order
\cite{AKLT} and for optical lattice implementations \cite{GarciaRipoll04}.
At the AKLT point $\beta=1/3$, the GS is given by a valence bond
solid (VBS) \cite{AKLT}, where each spin 1 is represented by a couple
of spin 1/2, provided the antisymmetric state is projected out. The
VBS state is constructed by forming short-ranged singlets between
nearest neighbor $S=1/2$ states and then symmetrizing local pairs
to get back $S=1$ states. For OBC there remains free effective spin-1/2
particles at the endpoints responsible for a fourfold degeneracy,
in an analogous way as in the the model (\ref{eq:alpha_delta_model})
with $\delta=1$ and $\alpha=0$. Away from $\beta=1/3$ the degeneracy
is lifted, the GS is a total singlet $S_{tot}=0$ and other valence
bond configurations give contribution to the GS. Anyway, the VBS state
is still a good approximation for a wide range of $\beta$'s, in particular
at the Heisenberg point $\beta=0$. Due to strong correlations in
the bulk, the two $S=1/2$ end spins tend to organize as a $\left|\Psi^{-}\right\rangle $
Bell state, in order to give rise to a total singlet. 

\begin{figure}
\includegraphics[%
  scale=0.4]{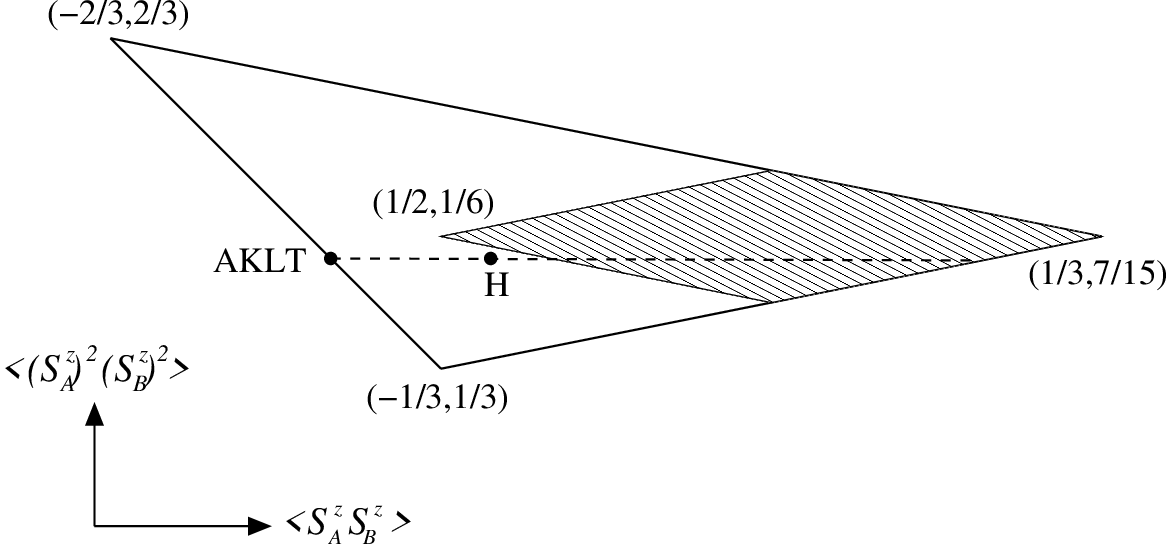}

\caption{Entangled (white) and separable (shaded) states of two globally SU(2)-invariant
qutrits are completely determined by means of $\langle S_{\textrm{A}}^{z}S_{\textrm{B}}^{z}\rangle$
and $\langle\left(S_{\textrm{A}}^{z}\right)^{2}\left(S_{\textrm{B}}^{z}\right)^{2}\rangle$.
\label{cap:entangled area}}
\end{figure}

Two different measures were considered to quantify the entanglement
between two spin-1. The VBS picture suggests the definition of the
partial concurrence ($\mathcal{PC}$) as the amount of entanglement
between the spin-1/2 belonging to different spin-1 particles. One
advantage of the $\mathcal{PC}$ is that it depends only on the \emph{z}-\emph{z}
correlator: $\mathcal{PC}_{\textrm{A,B}}=2\max\left\{ 0,2\left|\eta_{\textrm{AB}}^{zz}\right|-\eta_{\textrm{AB}}^{zz}-1/4\right\} $,
with $\eta_{\textrm{AB}}^{zz}=\langle S_{\textrm{A}}^{z}S_{\textrm{B}}^{z}\rangle/4$.
The symmetrization procedure distributes the entanglement among the
four qubit, so that the maximal possible value of the $\mathcal{PC}$
is $1/2$. However, the $\mathcal{PC}$ may fail in detecting genuine
qutrit entanglement, which is generally hard to quantify. In fact,
for qutrit mixed states there is no simple expression for the entanglement
of formation nor a simple criterion for separability is known. Nevertheless,
for SU(2)-rotationally invariant states, a necessary and sufficient
condition for a state to be entangled is that of having positive negativity
\cite{breuer05}, defined as\[
\mathcal{N}\left(\rho_{\textrm{AB}}\right)=\left\Vert \rho_{\textrm{AB}}^{T_{\textrm{A}}}\right\Vert _{1}-1,\]
where $\rho_{\textrm{AB}}^{T_{\textrm{A}}}$ stands for the partial
transpose with respect to subsystem A and $\left\Vert G\right\Vert _{1}=\tr\sqrt{GG^{\dagger}}$.
Calculating the negativity of a general SU(2)-invariant state, parametrized
by the quantities $\langle S_{\textrm{A}}^{z}S_{\textrm{B}}^{z}\rangle$
and $\langle\left(S_{\textrm{A}}^{z}\right)^{2}\left(S_{\textrm{B}}^{z}\right)^{2}\rangle$,
we are able to recognize the separable states. All the possible states
fall inside the triangle in Fig.~\ref{cap:entangled area}, whereas
the shaded area represents the separable states. 

In the AKLT case $\beta=1/3$, we choose the singlet among the four
degenerate states, because this is the state one would approach by
letting $\beta\rightarrow1/3$ and corresponds to the GS of the periodic
chain. From the exact solution, one finds $\langle S_{1}^{z}S_{L}^{z}\rangle=-\langle\left(S_{1}^{z}\right)^{2}\left(S_{L}^{z}\right)^{2}\rangle\simeq-4/9\left[1+6\left(-1\right)^{L}e^{-L/\xi_{\textrm{AKLT}}}\right]$,
where $\xi_{\textrm{AKLT}}=1/\ln\left(3\right)$ is the bulk AKLT
correlation length. It follows that in the thermodynamic limit $\mathcal{PC}=1/6$
and $\mathcal{N}=2/9$, where both values are approached exponentially
fast. This confirms the hypothesis that we have qubit as well as qutrit
entanglement.

At the Heisenberg point $\beta=0$ with OBC it is well established
the presence of surface order $\lim_{L\rightarrow\infty}\langle S_{1}^{z}S_{L}^{z}\rangle=-0.28306484(1)$
\cite{fath06} approached also in this case exponentially fast with
a bulk correlation length $\xi_{H}\sim6$. With accurate DMRG simulation
up to 100 sites we could establish a similar behavior for the correlations
$\langle\left(S_{1}^{z}\right)^{2}\left(S_{L}^{z}\right)^{2}\rangle$
with asymptotic value very close to $4/9$. These data imply the existence
of LDE in the Heisenberg model detected by a nonzero negativity $\mathcal{N}=0.0608426$,
even if qubit entanglement vanishes, i.e.~$\mathcal{PC}=0$. We note
here that both the Heisenberg (H) and the AKLT points (see Fig.~\ref{cap:entangled area})
lie on the line where $\langle\left(S_{1}^{z}\right)^{2}\left(S_{L}^{z}\right)^{2}\rangle\stackrel{L\rightarrow\infty}{\longrightarrow}\langle\left(S_{1}^{z}\right)^{2}\rangle\langle\left(S_{L}^{z}\right)^{2}\rangle=4/9$,
which means that the nonzero spins (effective charges) are uncorrelated.
Further enhancement of the LDE may be achieved in spin-1 models that
present also end-to-end charge correlations.

\paragraph{$S=1/2$ \emph{}Heisenberg model with probes. }

So far, we have considered situations where the probes are located
at the end points of a chain. Now we consider a different case: a
Heisenberg chain of length $L$ and two additional $S=1/2$ probes
$\vec{\tau}_{\textrm{A}}$ and $\vec{\tau}_{\textrm{B}}$ \[
\mathcal{H}=\sum_{j=1}^{L}\vec{\sigma}_{j}\cdot\vec{\sigma}_{j+1}+J_{p}\left(\vec{\sigma}_{1}\cdot\vec{\tau}_{\textrm{A}}+\vec{\sigma}_{d+1}\cdot\vec{\tau}_{\textrm{B}}\right)\]
 where $\vec{\tau}$ is a vector of Pauli matrices. The spin-probe
A interacts with the site 1, while B is connected to the site $d+1$.
The correlations between A and B will depend only by their distance
$d+2$, having assumed periodic boundary conditions (PBC) with $\sigma_{L+1}^{\alpha}\equiv\sigma_{1}^{\alpha}$. 

\begin{figure}
\begin{center}\includegraphics[%
  scale=0.32]{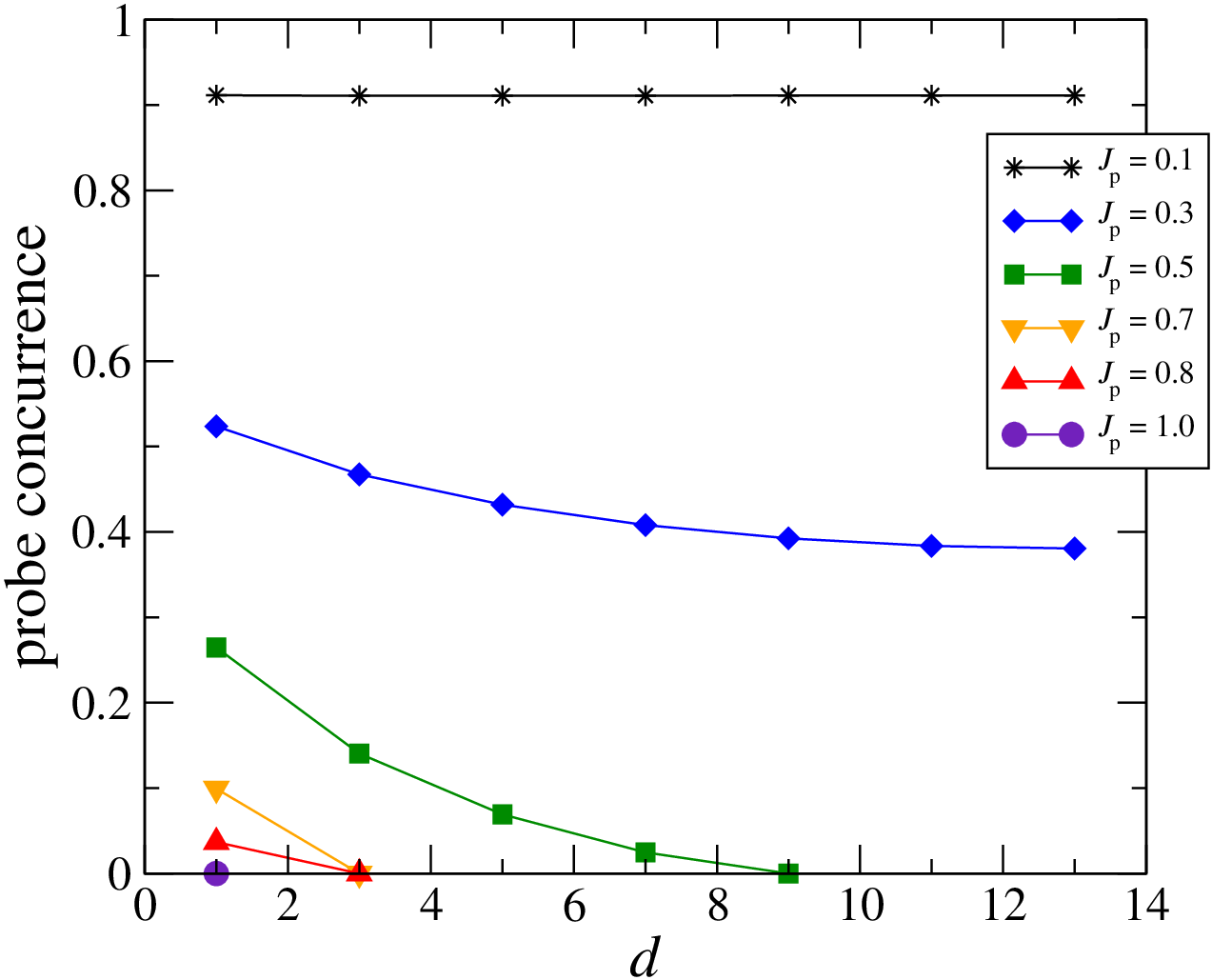}\end{center}

\caption{Concurrence between probes attached to a Heisenberg chain of length
$L=26$ as a function of the distance. The calculation was done for
various values of $J_{p}$. A dramatic increase of entanglement between
distant probes appears as $J_{p}$ is lowered. \label{cap:Concurrence_probes}}
\end{figure}

By means of exact diagonalization, we have studied the concurrence
$\mathcal{C}_{\textrm{AB}}(d)$, varying the distance at fixed $L$.
The results are illustrated in Fig.~\ref{cap:Concurrence_probes}
for a closed chain of $L=26$ (plus 2 probes). We have rejected even
values of $d$, as in these cases the GS is threefold degenerate,
belonging to the sector $S_{tot}=1$. Moreover, due to the PBC's the
maximum distance is reached at half chain, $d=13$. When $J_{p}=1$,
no probe entanglement is found, since $\mathcal{C}_{\textrm{AB}}(d)=0$
for every $d$. As we expected according to our considerations above,
$\mathcal{C}_{\textrm{AB}}$ is enhanced by weakening the interactions
between the probes and the spin chain. Already for $J_{p}=0.3$, the
entanglement is nonzero for every (odd) value of $d$ and remarkably
at $J_{p}=0.1$ the probes are almost completely entangled. Finite
size scaling of the concurrence between maximally distant probes,
$\mathcal{C}_{\textrm{AB}}(d=L/2,L)$ exhibits a slow decrease of
the concurrence with $L$ and it remains an open question whether
it survives at the thermodynamic limit. 

In addition, we just mention that changing the sign of the probe interactions
to ferromagnetic, $J_{p}<0$, the concurrence increases further, extending
the possibility of tailoring interactions that yield efficient entanglement
creation at large distance.  A similar behavior is observed by placing
the probes at the ends of an open chain. As above, the finite-size
effects are non-negligible due to the critical nature of the bulk.
Preliminary DMRG calculations with $L$ up to 100 leave open the possibility
of having LDE for ferromagnetic probe interactions. 

\paragraph{Conclusions. \emph{}}

With this Letter, we aim to bring to the attention of the QI community
a large class of spin-1/2 and spin-1 models capable of creating entanglement
between distant parties. On the one hand, this property opens up the
possibility to engineer QI devices like entanglers and quantum channels
using strongly correlated low-dimensional systems. In particular,
the phenomenon of concentrating the entanglement on the border of
finite-size system seems to be particularly suited for optical lattice
simulations. On the other hand, we observe that the transition point
where genuinely quantum correlations, signaled by the concurrence,
extend to long distance does not coincide with known QPT's. How this
issue embodies in the statistical mechanics framework is a challenging
question. Conversely, local measures of entanglement show a singular
behavior at QPT's that comes from the most relevant operator \cite{campos06}.
Specifically, in this work we have considered models with SU(2) symmetry,
which is common in nature and help to make the calculations easier.
Nonetheless, we verified that the results regarding LDE apply also
to non SU(2)-symmetric cases. Further work is in progress in order
to extend the investigation of LDE on other models, including electronic
systems. 

Many thanks to H. Nishimori for providing the Lanczos code TITPACK.
We are grateful to E. Ercolessi, F. Ortolani, M. Paris and S. Pasini
for useful discussions and especially to G. Morandi for a careful
reading of the manuscript. This work was supported by the TMR network
EUCLID (No.~HPRN-CT-2002-00325), and the COFIN projects 2002024522\_001
and 2003029498\_013.  \bibliographystyle{apsrev}
\bibliography{/home/marco/paperi/EPR_in_chains/entang}

\end{document}